# Polariton Waveguide Modes in Two-Dimensional Van der Waals Crystals: An Analytical Model and Correlative Scanning Near-Field Optical Microscopy Studies

*Fengsheng Sun,*[†,‡,§] *Wuchao Huang,*[†,‡,§] *Zebo Zheng,*[†,‡,§] *Ningsheng Xu,*[†,‡] *Yanlin Ke,*[†,‡] *Runze Zhan,*[†,‡] *Huanjun Chen,**[,†,‡] *and Shaozhi Deng**[,†,‡]

[†]State Key Laboratory of Optoelectronic Materials and Technologies, Guangdong Province Key Laboratory of Display Material and Technology, Sun Yat-sen University, Guangzhou 510275, China.

[‡]School of Electronics and Information Technology, Sun Yat-sen University, Guangzhou 510006, China.

*Corresponding authors: chenhj8@mail.sysu.edu.cn; stsdsz@mail.sysu.edu.cn.

§These authors contributed equally to the work.

**ABSTRACT** Two-dimensional van der Waals (vdW) crystals can sustain various types of polaritons with strong electromagnetic confinements, making them highly attractive for the nanoscale photonic and optoelectronic applications. While extensive experimental and numerical studies are devoted to the polaritons of the vdW crystals, analytical models are sparse. Particularly, applying such a model to describe the polariton behaviors visualized by state-of-art near-field optical microscopy requires further investigation. Herein, we develop an analytical waveguide model to describe the polariton propagations in vdW crystals. The dispersion contours, dispersion relations, and electromagnetic field distributions of different polariton waveguide modes are derived. The model is verified by near-field optical imaging and numerical simulation of phonon polaritons in the α-$MoO_3$, a typical vdW biaxial crystals. The model can be extended to other types of polaritons in vdW crystals, thus allowing for describing and understanding their localized electromagnetic behaviors analytically.

Polaritons, quasi-particles formed by strong coupling of electromagnetic waves with polarized charges in solids, can confine the free-space light field deeply down to the subwavelength scale. They have enabled various fascinating photonic effects and applications, such as negative refraction,[1,2] subdiffractional light focusing and guiding,[3–5] enhanced nonlinear optical effects,[6,7] to name but a few. Noble metals, highly-doped semiconductors, and bulk polar crystals are three representative polaritonic materials that are mostly studied in the past decades.[5–7] In recent years, the discovery of two-dimensional (2D) van der Waals (vdW) crystals represented by graphene have added a wealth of members to the polaritonic materials family,[8,9] such as graphene, hBN and α-$MoO_3$, and transition metal dichagenides, which respectively support plasmon polaritons,[10–14] phonon polaritons (PhP),[3–5,15–20] and exciton polaritons.[21–24] Due to the reduced dielectric screening in the 2D vdW crystals, their polaritons usually exhibit low loss with simultaneously strong electromagnetic confinement.[8–20] In addition, the polaritonic properties of the 2D crystals can be actively controlled *via* external stimulus, such as electrical gating and chemical doping.[13,14,19,25,26] More importantly, the vdW crystals can host polariton modes in the mid-infrared to terahertz spectral regions, where most of the conventional polaritonic materials cannot access.[8,9,17] These merits render the polaritonic vdW crystals highly attractive for the nanoscale photonics and optoelectronic applications.

From experimental aspects, far-field techniques, such as reflection, scattering, transmission, and near-field optical spectroscopies are usually employed to investigate the polaritons in the 2D vdW crystals. Particularly, the near-field approaches allow for revealing the intrinsic localized polaritonic field distributions and dispersion contours, which otherwise cannot be explored with the far-field techniques. However, in comparison with the extensive experimental studies, theoretical models describing the polariton modes are limited. Multilayer Fresenel reflection[12,15,19,20] and random phase approximation calculations[27] are the two most used analytical methods. They can retrieve the experimental polariton dispersion relations very well, while leaving the near-field polaritonic features inaccessible. Numerical methods, such as finite-difference time-domain and finite element (FEM) simulations, are able to calculate the particulars of the localized electromagnetic phenomena,[20,28] they however usually require massive calculations to reveal the dependence of the polaritonic behaviors on various parameters, e.g. excitation frequency, geometries of the system, etc. These are usually time-consuming and require costly computing resources. In particular, due to the low-symmetric crystalline structures of the vdW crystals, their permittivities are usually anisotropic, which will make the simulations much more complex. Additionally, the results obtained from the numerical methods often cannot reflect the clear physical pictures. Efforts have been devoted to establish analytical models. For example, an electromagnetic model treating the vdW crystals as pure 2D sheets is developed,[18] which is suitable for studying the polaritons in monolayer or few-layer vdW crystals of atomic-scale thicknesses. However, for vdW crystals with finite thicknesses, their polaritons are actually waveguide modes. The thicknesses of the crystals need to be considered for precisely calculating the polaritonic characteristics. To sum up, an analytical model capable of calculating the far-field and near-field characteristics of the polariton waveguide modes in the vdW crystals with anisotropic electromagnetic responses is in great demand. Two recent studies have proposed analytical solutions to calculate respectively the dispersions and electromagnetic modes in dielectric slabs of anisotropic permittivities,[29,30] while correlative experimental verifications, especially those from near-field optical measurements, are still lacking.

In this study, we developed an analytical model to calculate the electromagnetic behaviors of the polaritons hosted by vdW crystals of finite thicknesses. The localized electromagnetic field distributions, dispersion relations, and dispersion contours of the polariton waveguide modes can be obtained. The theoretical model is then applied to describe the PhP modes in a typical vdW crystals, the α-$MoO_3$, which is recently revealed as biaxial crystals sustaining natural hyperbolic responses. The results agree well with both of the correlative experimental near-field optical measurements and FEM simulations. Our model can therefore provide an analytical rationale to analyze and understand the experimental results on polaritons in the vdW crystals.

The vdW crystal is modeled as a 2D infinite waveguide of thickness *d*. (Figure 1). It is sandwiched between two semi-infinite plates, which act as the substrate and cover layer, respectively. The electromagnetic modes and the associated dispersion relations are obtained by solving the Maxwell's equations upon the continuities of the electric and magnetic fields at the two interfaces. The Cartesian coordinate system is set according to the three principal axes of the crystal, with the origin set at the interface between the 2D crystal and cover (Figure 1). Accordingly, the anisotropic permittivity tensor of the vdW crystal is expressed as,

$$\vec{\varepsilon}=\begin{pmatrix}\varepsilon_x & 0 & 0\\ 0 & \varepsilon_y & 0\\ 0 & 0 & \varepsilon_z\end{pmatrix} \tag{1}$$

where $\varepsilon_x$, $\varepsilon_y$, and $\varepsilon_z$ are permittivity components along the $x$, $y$, and $z$ axis, respectively, which are usually frequency-dependent. The substrate and cover layer are assumed to be isotropic, with permittivity as $\varepsilon_s$ and $\varepsilon_c$, respectively. Existence of polariton modes required that $\varepsilon_c$, $\varepsilon_s > 0$, and at least one of the components of ($\varepsilon_x$, $\varepsilon_y$, $\varepsilon_z$) is negative. For isotropic ($\varepsilon_x=\varepsilon_y=\varepsilon_z$) and uniaxial ($\varepsilon_x=\varepsilon_y\neq\varepsilon_z$) crystals, Maxwell's equations have two sets of independent solutions. They are respectively transverse magnetic modes (TM: $E_x$, $H_y$, $E_z$)[9,14] and transverse electric modes (TE: $H_x$, $E_y$, $H_z$) (Note 1, Supporting Information). However, in biaxial crystals these two solutions are no longer independent. An analytical model describing the complex polariton modes in the biaxial crystals should therefore consider the polarization mixing effects. For that reason, we consider a polariton waves propagating in the basal plane ($x$–$y$ plane) of the crystal along $\theta$ with respect to $x$-axis. We first introduce a coordinate transformation,

$$X'=\begin{pmatrix}x'\\ y'\\ z'\end{pmatrix}=\hat{T}X=\begin{pmatrix}\cos\theta & \sin\theta & 0\\ -\sin\theta & \cos\theta & 0\\ 0 & 0 & 1\end{pmatrix}\begin{pmatrix}x\\ y\\ z\end{pmatrix} \tag{1}$$

The effective permittivity tensor $\vec{\varepsilon}'$ is then expressed as,

$$\vec{\varepsilon}'=\hat{T}\vec{\varepsilon}\hat{T}^{-1}=\begin{pmatrix}\varepsilon_{xx} & \varepsilon_{xy} & 0\\ \varepsilon_{yx} & \varepsilon_{yy} & 0\\ 0 & 0 & \varepsilon_{zz}\end{pmatrix}=\begin{pmatrix}\varepsilon_x\cos^2\theta+\varepsilon_y\sin^2\theta & \left(\varepsilon_y-\varepsilon_x\right)\sin\theta\cos\theta & 0\\ \left(\varepsilon_y-\varepsilon_x\right)\sin\theta\cos\theta & \varepsilon_y\cos^2\theta+\varepsilon_x\sin^2\theta & 0\\ 0 & 0 & \varepsilon_z\end{pmatrix} \tag{2}$$

A typical polariton mode can be expressed as $\vec{E}\left(x',z',t\right)=\vec{e}E\left(z'\right)\exp(iqx'-i\omega t)$, where $\vec{e}$ is the unit vector of the electric field and $q$ is the wave vector along $x'$ axis. We then arrive at the following equations for the electric fields,

$$\nabla^2\vec{E}+k_0^2\vec{\varepsilon}'\vec{E}=\nabla(\nabla\cdot\vec{E}) \tag{3}$$

where $k_0=\sqrt{\omega^2\mu_0\varepsilon_0}=2\pi/\lambda_0$ is the free-space wave vector. The general solutions of Eq. (3) should have the form as,

$$\vec{E}^{(\pm)}\left(x',z'\right)=\vec{A}^{(\pm)}\exp(\pm ik_z z')\exp(iqx') \tag{4}$$

where $\vec{A}^{(\pm)}$ and $\pm k_z$ are amplitudes and propagation constants of the polariton waves propagating towards +*z* and –*z* directions, respectively. Substituting Eq. (4) into Eq. (3), the following equations of the electric fields $\vec{E}^{(\pm)}$ can be obtained,

$$M^{(a)}\cdot\vec{E}^{(\pm)}=\begin{pmatrix} k_0^2\varepsilon_{xx}-k_z^2 & k_0^2\varepsilon_{xy} & \pm qk_z \\ k_0^2\varepsilon_{yx} & k_0^2\varepsilon_{yy}-k_z^2-q^2 & 0 \\ \pm qk_z & 0 & k_0^2\varepsilon_{zz}-q^2 \end{pmatrix}\begin{pmatrix} E_x^{(\pm)} \\ E_y^{(\pm)} \\ E_z^{(\pm)} \end{pmatrix}=0 \tag{5}$$

To have non-trivial solutions, it is required that det[$M^{(a)}$] = 0, which results in,

$$k_z^4\varepsilon_{zz}+k_z^2[q^2(\varepsilon_{xx}+\varepsilon_{zz})-k_0^2(\varepsilon_{xx}+\varepsilon_{yy})\varepsilon_{zz}]+q^4\varepsilon_{xx}-q^2k_0^2[\varepsilon_{xx}(\varepsilon_{yy}+\varepsilon_{zz})-\varepsilon_{xy}\varepsilon_{yx}]+k_0^4\varepsilon_{zz}(\varepsilon_{xx}\varepsilon_{yy}-\varepsilon_{xy}\varepsilon_{yx})=0 \tag{6}$$

The solutions then read as,

$$k_z^2=\frac{1}{2\varepsilon_{zz}}\left\{\left[k_0^2(\varepsilon_{xx}+\varepsilon_{yy})\varepsilon_{zz}-q^2(\varepsilon_{xx}+\varepsilon_{zz})\right]\pm\sqrt{\Delta}\right\} \tag{7}$$

with parameter Δ expressed as,

$$\Delta=\sqrt{[q^2(\varepsilon_{xx}+\varepsilon_{zz})-k_0^2(\varepsilon_{xx}+\varepsilon_{yy})\varepsilon_{zz}]^2-4\varepsilon_{zz}\left\{q^4\varepsilon_{xx}-q^2k_0^2[\varepsilon_{xx}(\varepsilon_{yy}+\varepsilon_{zz})-\varepsilon_{xy}\varepsilon_{yx}]+k_0^4\varepsilon_{zz}(\varepsilon_{xx}\varepsilon_{yy}-\varepsilon_{xy}\varepsilon_{yx})\right\}} \tag{8}$$

On the other hand, in an isotropic medium, the $\vec{E}^{(\pm)}$ fulfill,

$$M^{(i)}\cdot\vec{E}^{(\pm)}=\begin{pmatrix} k_0^2\varepsilon_{c,s}-k_z^2 & 0 & \pm qk_z \\ 0 & k_0^2\varepsilon_{c,s}-k_z^2-q^2 & 0 \\ \pm qk_z & 0 & k_0^2\varepsilon_{c,s}-q^2 \end{pmatrix}\begin{pmatrix} E_x^{(\pm)} \\ E_y^{(\pm)} \\ E_z^{(\pm)} \end{pmatrix}=0 \tag{9}$$

and $(k_0^2\varepsilon_{c,s}-k_z^2-q^2)^2=0$ (10)

For a vdW crystal of finite thickness (Figure 1), the electromagnetic fields in the cover layer (*c*), waveguide layer (vdW crystal, *w*), and substrate layer (*s*) can be expressed as,

$$\vec{E}\ or\ \vec{H}(z',x') = \begin{cases} \vec{A}_{E,H}^{(c)} \exp(-\alpha_z^{(c)} z') \exp(iqx'), \quad z' \geq 0 \\ [\vec{A}_{E,H}^{(w,+)} \exp(ik_z^{(w)} z') + \vec{A}_{E,H}^{(w,-)} \exp(-ik_z^{(w)} z')] \exp(iqx'), \quad -d < z' < 0 \\ \vec{A}_{E,H}^{(s)} \exp(\alpha_z^{(s)} z') \exp(iqx'), \quad z' \leq -d \end{cases} \tag{11}$$

where $\alpha_z^{(c,s)} = -ik_z^{(c,s)} = \sqrt{q^2 - k_0^2 \varepsilon_{c,s}}$ . The amplitude coefficients are,

$$\begin{cases} \vec{A}_E^{(w,\pm)} = (A_x^{(w,\pm)}, W_{y/x}^{(\pm)} A_x^{(w,\pm)}, W_{z/x}^{(\pm)} A_x^{(w,\pm)})^T, \quad \vec{A}_H^{(w,\pm)} = \dfrac{i}{\omega\mu_0} (A_x^{(w,\pm)}, W_{y/x}^{(\pm)} A_x^{(w,\pm)}, W_{z/x}^{(\pm)} A_x^{(w,\pm)})^T, \\ \vec{A}_E^{(c)} = (A_{x,p}^{(c)}, A_{y,s}^{(c)}, C_{z/x} A_{x,p}^{(c)})^T, \quad \vec{A}_H^{(c)} = \dfrac{i}{\omega\mu_0} (\alpha_z^{(c)} A_{y,s}^{(c)}, (-\alpha_z^{(c)} - iqC_{z/x}) A_{x,p}^{(c)}, iqA_{y,s}^{(c)})^T, \\ \vec{A}_E^{(s)} = (A_{x,p}^{(s)}, A_{y,s}^{(s)}, S_{z/x} A_{x,p}^{(s)})^T, \quad \vec{A}_H^{(s)} = \dfrac{i}{\omega\mu_0} (\alpha_z^{(s)} A_{y,s}^{(s)}, (\alpha_z^{(s)} - iqS_{z/x}) A_{x,p}^{(s)}, iqA_{y,s}^{(s)})^T. \end{cases} \tag{12}$$

with

$$C_{z/x} = \frac{-iq\alpha_z^{(c)}}{k_0^2 \varepsilon_c - q^2}, \quad W_{y/x}^{(\pm)} = \frac{-k_0^2 \varepsilon_{yx}}{k_0^2 \varepsilon_{yy} - k_z^{(w)2} - q^2}, \quad W_{z/x}^{(\pm)} = \frac{\pm qk_z^{(w)}}{k_0^2 \varepsilon_{zz} - q^2}, \quad S_{z/x} = \frac{iq\alpha_z^{(s)}}{k_0^2 \varepsilon_s - q^2} \tag{13}$$

The solutions of the electric and magnetic fields can be determined by matching the boundary conditions that tangential components of $\vec{E}$ and $\vec{H}$ should be continuous at the two interfaces ($z'$ = 0 and $z' = -d$),

$$\begin{cases} E_{x,y}^{(c)} = E_{x,y}^{(w)}, \quad H_{x,y}^{(c)} = H_{x,y}^{(w)} \quad (z' = 0) \\ E_{x,y}^{(w)} = E_{x,y}^{(s)}, \quad H_{x,y}^{(w)} = H_{x,y}^{(s)} \quad (z' = -d) \end{cases} \tag{14}$$

Therefore, the amplitude coefficients should fulfill $M \cdot (A_{x,p}^{(c)}, A_x^{(w,+)}, A_x^{(w,-)}, A_{x,p}^{(s)})^T = 0$, where,

$$M = \begin{pmatrix} 1 & -1 & -1 & 0 \\ -\alpha_z^{(c)} - iqC_{z/x} & -ik_z^{(w)} + iqW_{z/x}^{(+)} & ik_z^{(w)} + iqW_{z/x}^{(-)} & 0 \\ 0 & \exp(-ik_z^{(w)} d) & \exp(ik_z^{(w)} d) & \exp(-\alpha_s d) \\ 0 & (-ik_z^{(w)} + iqW_{z/x}^{(+)}) \exp(-ik_z^{(w)} d) & (ik_z^{(w)} + iqW_{z/x}^{(-)}) \exp(ik_z^{(w)} d) & (\alpha_z^{(s)} - iqS_{z/x}) \exp(-\alpha_z^{(s)} d) \end{pmatrix} \tag{15}$$

The polariton waveguide modes and the associated dispersion relations can thereafter be obtained by solving the equation det($M$) = 0. Specifically, the dispersion relation can be stated as,

$$\frac{[(\alpha_z^{(c)}+iqC_{z/x})+(ik_z^{(w)}-iqW_{z/x}^{(+)})][(\alpha_z^{(s)}-iqS_{z/x})+(ik_z^{(w)}+iqW_{z/x}^{(-)})]\exp(2ik_z^{(w)}d)}{[(\alpha_z^{(c)}+iqC_{z/x})-(ik_z^{(w)}+iqW_{z/x}^{(+)})][(\alpha_z^{(s)}-iqS_{z/x})-(ik_z^{(w)}-iqW_{z/x}^{(-)})]}=1 \tag{16}$$

Eq. (16) clearly indicates that in-plane wave vectors of a specific waveguide mode is strongly dependent on the propagation direction $\theta$ and thickness of the vdW crystal. Once the dispersion relation has been determined, electric and magnetic fields of the polariton waveguide modes can be obtained according to Eqs. (12) and (13).

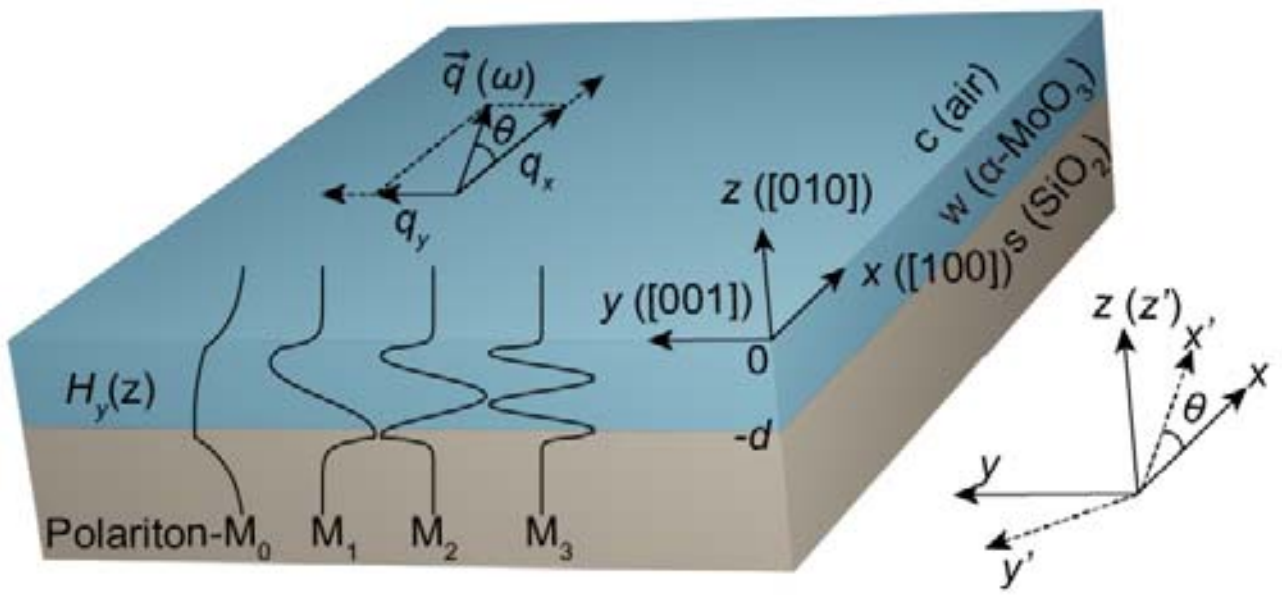

**Figure 1.** Schematic showing the polariton waveguide model. The distributions of the magnetic fields, $H_y(z)$, are analytical results from the model. Letters "*c*", "*w*", and "*s*" stand for cover, waveguide, and substrate layers, respectively. Right corner: schematic showing the rotation coordinate transform.

The model proposed above can be extended to describe far-field responses, *i.e.*, reflection spectroscopy, of the polaritonic vdW crystal. To that end, Fresnel coefficients were calculated by including the incident, reflected, and transmitted electromagnetic fields. Specifically, the electric and magnetic field distributed in the multilayer structure can be expressed as,

$$
\vec{E}\ or\ \vec{H}(z',x')=\begin{cases}[\vec{A}_{E,H}^{(c,+)}\exp(ik_z^{(c)}z')+\vec{A}_{E,H}^{(c,-)}\exp(-ik_z^{(c)}z')]\exp(iqx'), & z'\geq 0\\ [\vec{A}_{E,H}^{(w,+)}\exp(ik_z^{(w)}z')+\vec{A}_{E,H}^{(w,-)}\exp(-ik_z^{(w)}z')]\exp(iqx'), & -d<z'<0\\ \vec{A}_{E,H}^{(s,-)}\exp(-ik_z^{(s)}z')\exp(iqx'), & z'\leq -d\end{cases} \tag{17}
$$

where

$$
\begin{cases}\vec{A}_E^{(c,\pm)}=(A_{x,p}^{(c,\pm)},A_{y,s}^{(c,\pm)},C_{z/x}^{(\pm)}A_{x,p}^{(c,\pm)})^T, \quad \vec{A}_H^{(c,\pm)}=\dfrac{i}{\omega\mu_0}(\mp ik_z^{(c)}A_{y,s}^{(c,\pm)},(\pm ik_z^{(c)}-iqC_{z/x}^{(\pm)})A_{x,p}^{(c,\pm)},iqA_{y,s}^{(c,\pm)})^T,\\ \vec{A}_E^{(s,-)}=(A_{x,p}^{(s,-)},A_{y,s}^{(s,-)},S_{z/x}^{(-)}A_{x,p}^{(s,-)})^T, \quad \vec{A}_H^{(s,-)}=\dfrac{i}{\omega\mu_0}(ik_z^{(s)}A_{y,s}^{(s,-)},(-ik_z^{(s)}-iqS_{z/x}^{(-)})A_{x,p}^{(s,-)},iqA_{y,s}^{(s,\pm)})^T.\end{cases}
$$

$$
C_{z/x}^{(\pm)}=\frac{\pm qk_z^{(c)}}{k_0^2\varepsilon_c-q^2}, \quad S_{z/x}^{(-)}=\frac{-qk_z^{(s)}}{k_0^2\varepsilon_s-q^2}. \tag{18}
$$

By matching the boundary conditions in Eq. (14), the following relations between different amplitude coefficients can be obtained,

$$
\begin{cases}A_x^{(w,+)}=\dfrac{(k_z^{(w)}+qW_{z/x}^{(-)})-(k_z^{(s)}+qS_{z/x}^{(-)})}{(k_z^{(w)}+qW_{z/x}^{(-)})+(k_z^{(w)}-qW_{z/x}^{(+)})}\exp[i(k_z^{(s)}+k_z^{(w)})d]A_{x,p}^{(s,-)}\\ A_x^{(w,-)}=\dfrac{(k_z^{(w)}-qW_{z/x}^{(+)})+(k_z^{(s)}+qS_{z/x}^{(-)})}{(k_z^{(w)}+qW_{z/x}^{(-)})+(k_z^{(w)}-qW_{z/x}^{(+)})}\exp[i(k_z^{(s)}-k_z^{(w)})d]A_{x,p}^{(s,-)}\\ A_x^{(w,+)}=\dfrac{k_z^{(w)}-k_z^{(s)}}{2k_z^{(w)}W_{y/x}^{(+)}}\exp[i(k_z^{(s)}+k_z^{(w)})d]A_{y,s}^{(s,-)}\\ A_x^{(w,-)}=\dfrac{k_z^{(w)}+k_z^{(s)}}{2k_z^{(w)}W_{y/x}^{(-)}}\exp[i(k_z^{(s)}-k_z^{(w)})d]A_{y,s}^{(s,-)}\\ A_{x,p}^{(c,+)}=\dfrac{(k_z^{(c)}+qC_{z/x}^{(-)})+(k_z^{(w)}+qW_{z/x}^{(+)})}{(k_z^{(w)}-qC_{z/x}^{(+)})+(k_z^{(c)}+qC_{z/x}^{(-)})}A_x^{(w,+)}+\dfrac{(k_z^{(c)}+qC_{z/x}^{(-)})-(k_z^{(w)}+qW_{z/x}^{(-)})}{(k_z^{(w)}-qC_{z/x}^{(+)})+(k_z^{(c)}+qC_{z/x}^{(-)})}A_x^{(w,-)}\\ A_{x,p}^{(c,-)}=\dfrac{(k_z^{(c)}-qC_{z/x}^{(-)})-(k_z^{(w)}-qW_{z/x}^{(+)})}{(k_z^{(w)}-qC_{z/x}^{(+)})+(k_z^{(c)}+qC_{z/x}^{(-)})}A_x^{(w,+)}+\dfrac{(k_z^{(c)}-qC_{z/x}^{(-)})+(k_z^{(w)}+qW_{z/x}^{(-)})}{(k_z^{(w)}-qC_{z/x}^{(+)})+(k_z^{(c)}+qC_{z/x}^{(-)})}A_x^{(w,-)}\\ A_{y,s}^{(c,+)}=\dfrac{(k_z^{(c)}+k_z^{(w)})W_{y/x}^{(+)}}{2k_z^{(c)}}A_x^{(w,+)}+\dfrac{(k_z^{(c)}-k_z^{(w)})W_{y/x}^{(-)}}{2k_z^{(c)}}A_x^{(w,-)}\\ A_{y,s}^{(c,-)}=\dfrac{(k_z^{(c)}-k_z^{(w)})W_{y/x}^{(+)}}{2k_z^{(c)}}A_x^{(w,+)}+\dfrac{(k_z^{(c)}+k_z^{(w)})W_{y/x}^{(-)}}{2k_z^{(c)}}A_x^{(w,-)}\end{cases} \tag{19}
$$

Accordingly, Fresnel reflection and transmission coefficients can be expressed as,

$$
\begin{cases}
r_{ss} = \dfrac{A_{y,s}^{(c,+)}}{A_{y,s}^{(c,-)}}, \quad r_{ps} = \dfrac{A_{x,p}^{(c,+)}}{A_{y,s}^{(c,-)}}, \quad r_{sp} = \dfrac{A_{y,s}^{(c,+)}}{A_{x,p}^{(c,-)}}, \quad r_{pp} = \dfrac{A_{x,p}^{(c,+)}}{A_{x,p}^{(c,-)}}, \\
t_{ss} = \dfrac{A_{y,s}^{(s,-)}}{A_{y,s}^{(c,-)}}, \quad t_{ps} = \dfrac{A_{x,p}^{(s,-)}}{A_{y,s}^{(c,-)}}, \quad t_{sp} = \dfrac{A_{y,s}^{(s,-)}}{A_{x,p}^{(c,-)}}, \quad t_{pp} = \dfrac{A_{x,p}^{(s,-)}}{A_{x,p}^{(c,-)}}.
\end{cases} \tag{20}
$$

In order to demonstrate the effectiveness of the model, it is applied to calculate the polariton waveguide modes in a typical polar vdW crystal, α-$MoO_3$, which was recently revealed as a biaxial crystal exhibiting natural hyperbolic responses.[18–20,31] The unit cell of the α-$MoO_3$ is composed of three nonequivalent Mo–O bonds along the [100], [001], and [010] crystalline axes, respectively (Figure 2a).[32,33] The coordinate system is then set according to these three principal axes (Figure 2a, inset). The corresponding phonon modes result in three Reststrahlen bands in the mid-infrared range, which are Band 1 in the range of 545 $cm^{-1}$ to 851 $cm^{-1}$, and Bands 2 and 3 in the ranges of 820 $cm^{-1}$ to 972 $cm^{-1}$ and 958 $cm^{-1}$ to 1010 $cm^{-1}$, respectively.[18–20] The permittivities of the crystal in these three bands are governed by phonon transitions, which can be calculated using a Lorentzian model (Note 2, Supporting Information). Within each Reststrahlen band, the three principal permittivity components always exhibit real parts with opposite signs (Figure 2b), enabling the α-$MoO_3$ hyperbolic along different axes for each band. These can be seen more clearly by drawing the isofrequency surfaces corresponding to the three Reststrahlen bands, which are hyperboloids with two opening surfaces (Figure 2c, Note 2, Supporting Information). In particular, in Band 1 ($\mathrm{Re}(\varepsilon_y) < 0$ and $\mathrm{Re}(\varepsilon_x) \neq \mathrm{Re}(\varepsilon_z) > 0$) and Band 2 ($\mathrm{Re}(\varepsilon_x) < 0$ and $\mathrm{Re}(\varepsilon_y) \neq \mathrm{Re}(\varepsilon_z) > 0$), the in-plane dispersions are hyperbolic (Figure S1, Supporting Information), making the corresponding electromagnetic waves highly anisotropic (Figure S2, Supporting Information).

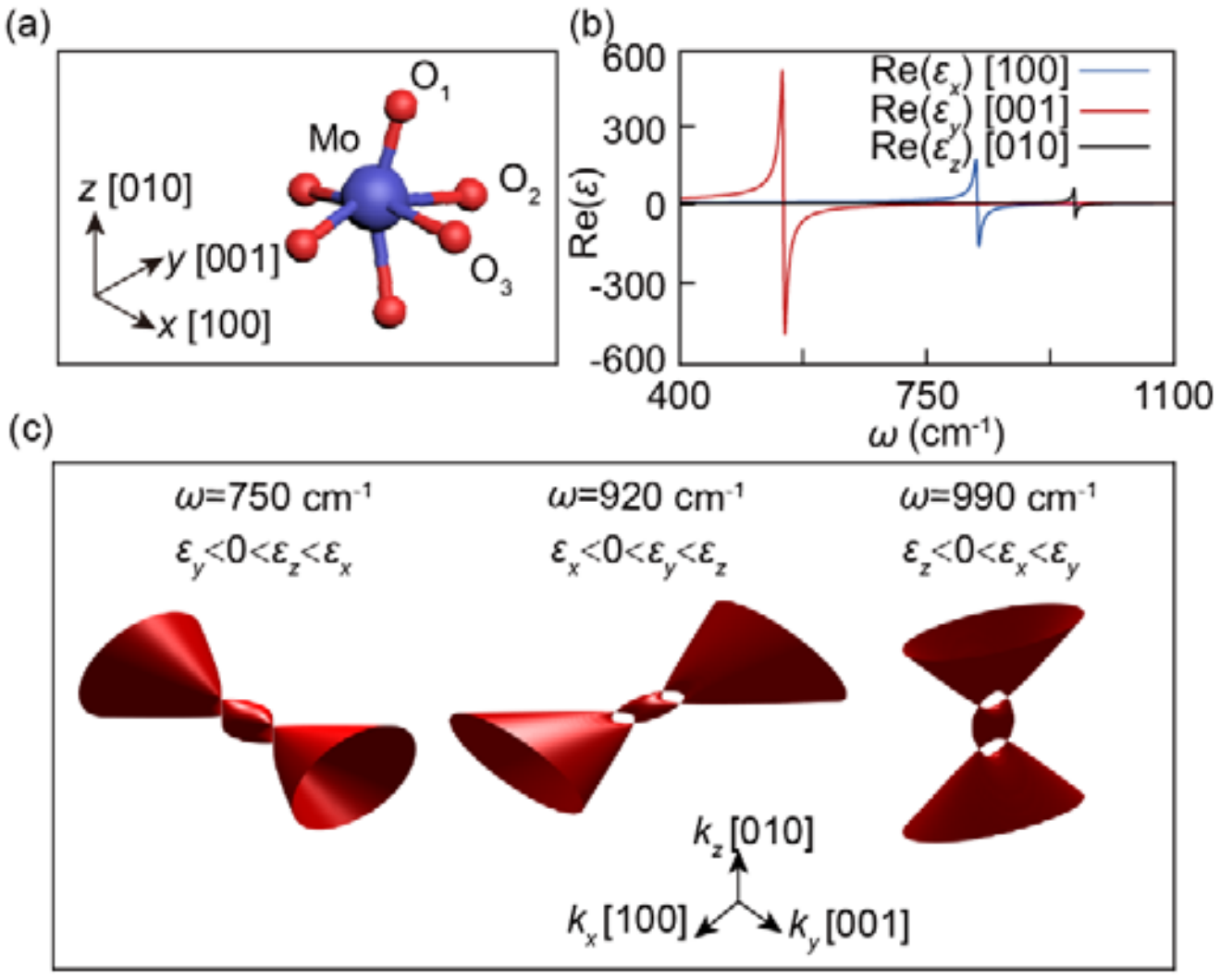

**Figure 2.** Dispersion of the biaxial α-$MoO_3$ crystal. (a) Schematic showing the unit cell of the α-$MoO_3$ crystal. The vibrations of the three covalent bonds, Mo–$O_1$, Mo–$O_2$, and Mo–$O_3$ are distinct with each other. Inset: Cartesian coordinate system in the α-$MoO_3$ crystal. (b) Real parts of the permittivities, Re($\varepsilon$), along the three principal axes. (c) Three-dimensional isofrequency surfaces of electromagnetic waves propagating inside the α-$MoO_3$ crystal.

PhPs are formed by the strong coupling of the phonons in the three Reststrahlen bands with electromagnetic waves. With the effective permittivity tensor and dispersion relation, which are respectively determined by Eqs. (3) and (15), one can calculate the frequency-dependent dispersion contours, $\omega(q_x, q_y)$, of the PhP modes hosted by the α-$MoO_3$ 2D waveguide. Figure 3a shows the results for a 2D flake of 210 nm thick, which is supported onto a $SiO_2$ substrate and surrounded by air. The complex dispersion contours are constructed by stacking the in-plane dispersion contours along the frequency axis (Figure 3a). For every frequency the in-plane 2D dispersion is highly anisotropic. Specifically, for the Reststrahlen Band 1 and 2, the in-plane permittivities with real parts of opposite signs generate PhP waveguide modes with in-plane hyperbolicity (Figure 3b and 3c, white lines). When the frequency is increased into the Band 3 with positive but unequal Re($\varepsilon_x$) and Re($\varepsilon_y$), the in-plane PhP dispersions evolve from hyperbola to ellipses (Figure 3d, white line).

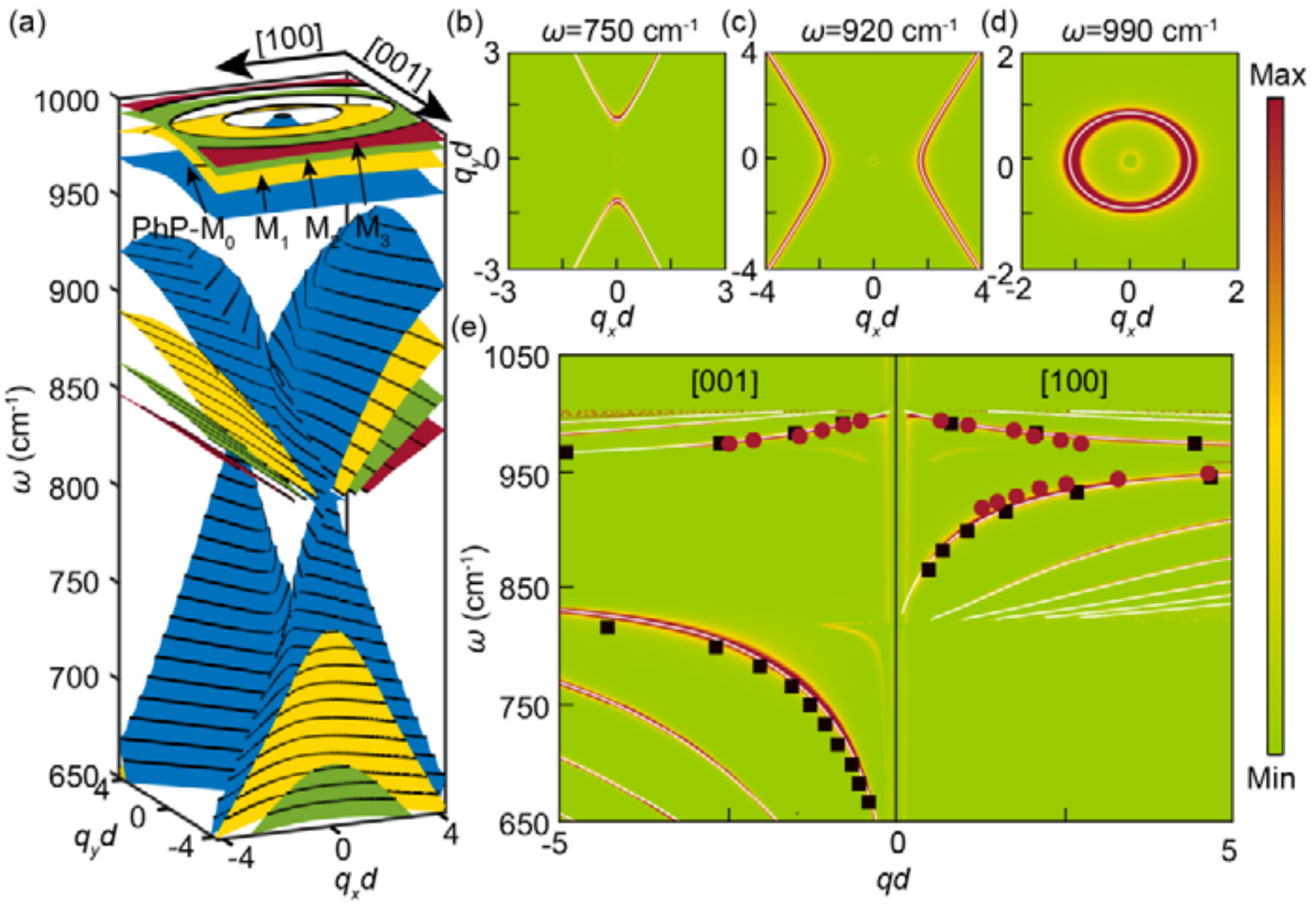

**Figure 3.** Dispersions of the PhPs in the α-$MoO_3$ 2D waveguide. (a) Frequency-dependent dispersion contours of the PhP modes. The colors delegate different order modes: PhP-$TM_0$ (blue), PhP-$TM_1$ (yellow), PhP-$TM_2$ (green), and PhP-$TM_3$ (red). (b)–(d) In-plane dispersion contours of the PhP modes excited at 750 $cm^{-1}$ (Band 1), 920 $cm^{-1}$ (Band 2), and 990 $cm^{-1}$ (Band 3), respectively. (e) Dispersion relations of the PhPs. The red dots indicate experimental data extracted from near-field optical microscopy. The black squares are the results obtained from finite element method simulations. The pseudocolored image represents the calculated imaginary part of the complex reflectivity of the air/a-$MoO_3$/$SiO_2$ multilayered structure. The white lines shown in (b)–(e) are results calculated by the waveguide model. The thickness of the α-$MoO_3$ flake is $d$ = 210 nm. All of the wave vectors are multiplied by the thickness of the α-$MoO_3$ flake.

The dispersion relation, $\omega(\bar{q})$, for the PhP can be derived by intersecting the isofrequency contour with a plane passing through its center along the direction of $\bar{q}$. Figure 3e plots two typical results for PhPs propagating along the [100] ($x$-axis) and [001] ($y$-axis) directions (white lines). Both of the Type I ($\mathrm{Re}(\varepsilon_x)\cdot\mathrm{Re}(\varepsilon_y) < 0$ and $\mathrm{Re}(\varepsilon_z) > 0$) and Type II ($\mathrm{Re}(\varepsilon_x, \varepsilon_y) > 0$ and $\mathrm{Re}(\varepsilon_z) < 0$) hyperbolic dispersions are observed. In addition, in Band 1 (Band 2) the PhP waves can only propagate along the $x$-axis ($y$-axis), while for those in Band 3, propagations along both directions are allowed (Figure 3e, white lines). These results clearly demonstrate the highly in-plane anisotropic behaviors of the PhP waves in the α-$MoO_3$.

It is noted that the multilayer Fresnel reflection coefficient calculations are widely used to describe the dispersion relations and contours of the polaritons in the 2D vdW crystals.[12,15,19,20] The results obtained from our model match very well with the Fresnel calculations (Figure 3b–3e, pseudocolored images and white lines). Furthermore, our model endows the origin of the PhPs as waveguide modes. Another interesting aspect is that high-order waveguide modes can be observed on both of the isofrequency contours and dispersion relations (Figure 3a and 3e). These modes exhibit larger in-plane wave vectors in comparison with the first-order mode ($l$ = 0: PhP-$M_0$). They therefore allow for stronger electromagnetic confinements.

The waveguide model is able to calculate the polariton field distributions in the α-$MoO_3$, which cannot be accessed using the far-field calculations (e.g. Fresnel reflection coefficient calculation and random phase approximation calculations). A representative example is given in Figure 1a at an excitation frequency of 990 $cm^{-1}$ (Band 3). For PhP propagating along the $x$-axis, the electromagnetic fields are strongly confined inside the α-$MoO_3$ layer, with evanescent components exponentially decaying into the air and substrate (Figure 1a). The number of nodes increases for PhP mode with a larger $l$, which is a typical characteristic for waveguide modes.

These features are similar for PhP waves with arbitrary in-plane wave vectors (Figure S2 and S3, Supporting Information).

In order to experimentally verify the effectiveness of the waveguide model, real-space nano-imaging was performed to visualize the PhP propagations inside the α-$MoO_3$. Specifically, α-$MoO_3$ flakes were synthesized and deposited onto the $SiO_2$ substrate (Figure 4a, see Methods for more details). The nano-imaging was conducted on a scattering-type scanning near-field optical microscope (s-SNOM).[34,35] A circular hole with diameter of 590 nm was lithographically drilled onto a α-$MoO_3$ flake with thickness of 210 nm (Figure 4b). In a specific measurement, the PhP waves will be launched by the metallic tip of the s-SNOM using a tunable mid-infrared laser. These waves will interfere with those reflected by the edge of the hole and generate interference fringes (Figure S4, Supporting Information), whereby the PhP characteristics, such as the wavelength and dispersion contours and relations, can be obtained by analyzing these fringes. Typical near-field images using excitation frequencies of 920 $cm^{-1}$ (Band 2) and 990 $cm^{-1}$ (Band 3) are shown in Figure 4c and 4d, which clearly indicate anisotropic PhP propagations. Specifically, for the excitation at 920 $cm^{-1}$, a wavefront of hyperbolic shape can be observed (Figure 4c), while it changes into an elliptical shape at the frequency of 990 $cm^{-1}$ (Figure 4d). Due to the finite curvature of the metal tip of the s-SNOM, only the first-order ($l = 0$) polariton modes can be visualized in the nano-imaging measurements. These results are consistent with the analytical model calculations (Figure 4g and 4h, Note 3, Supporting Information) and FEM simulations (Figure 4k and 4l, Figure S4a, Figure S5a, see Methods for details). The comparison of the PhP patterns can in principle be made in Band 1 with in-plane hyperbolicity (Figure 4f and 4j), which however is inaccessible because the excitation frequencies are beyond the operation frequency range of our s-SNOM system.

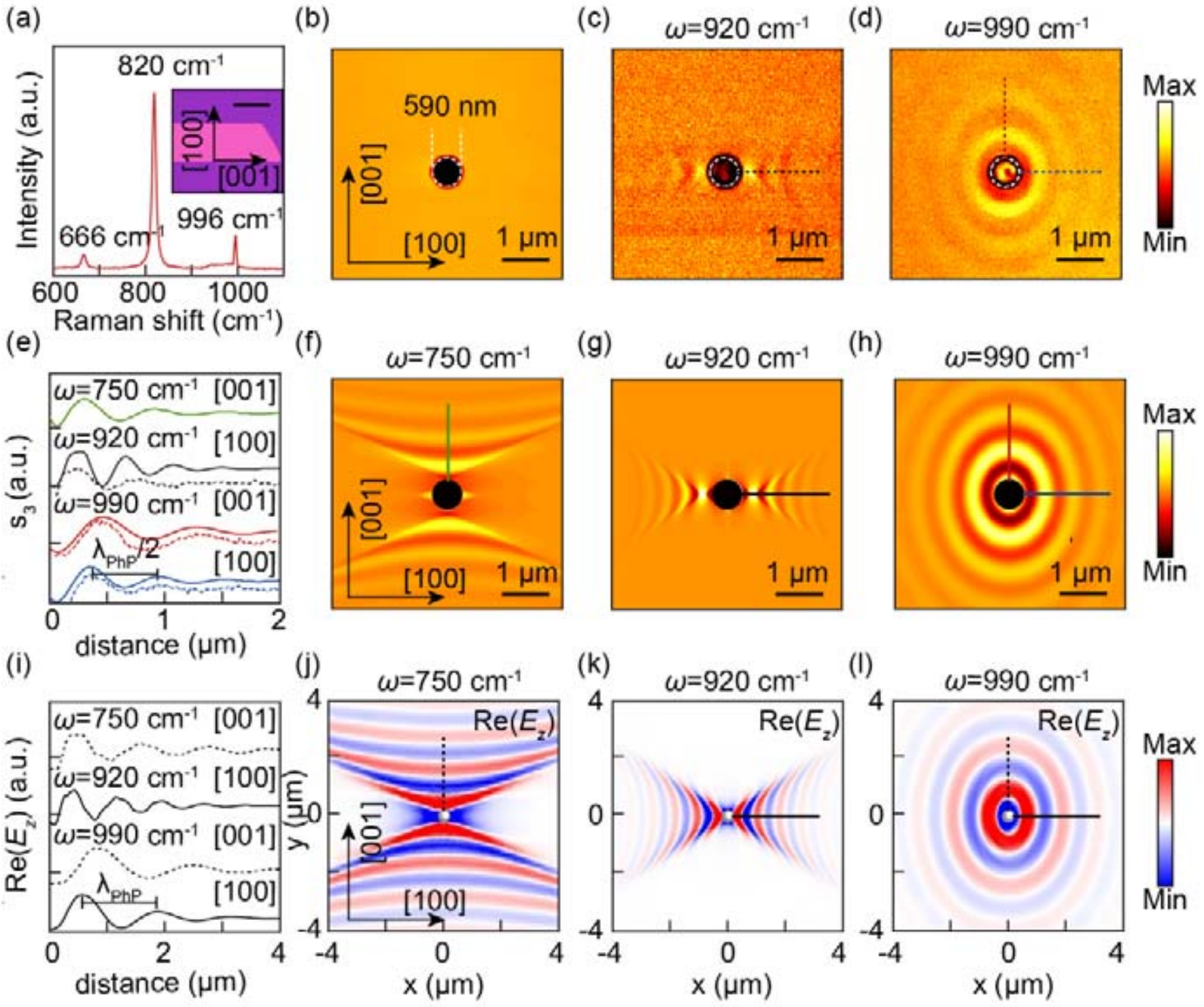

**Figure 4.** Verifications of the polariton waveguide model. (a) Raman spectrum of a typical α-$MoO_3$ flake. Inset: digital photograph of the flake. Scale bar is 200 μm. (b) Atomic force microscope topography of the circular hole in α-$MoO_3$ flake. The diameter of the hole is 590 nm and the thickness of the flake is 210 nm. (c, d) Near-filed optical intensity distributions of the α-$MoO_3$ flake. The excitation frequencies are (c) 920 $cm^{-1}$ (Band 2) and (d) 990 $cm^{-1}$ (Band 3), respectively. The circular holes are indicated by dashed white lines. (e) Comparison between the experimental (dashed lines) and theoretical (solid lines) near-field amplitude profiles. The profiles are extracted along the lines depicted in (c, d: dashed lines) and (f–h: solid lines). (f–h) Theoretical near-field optical intensity distributions of the PhP waves on the α-$MoO_3$ flake with a circular hole. The excitation frequencies are (f) 750 $cm^{-1}$ (Band 1), (g) 920 $cm^{-1}$ (Band 2), and (h) 990 $cm^{-1}$ (Band 3), respectively. (i) Amplitude profiles along the lines depicted in (j–l). (j)–(l) Calculated real parts of *z*-components, Re($E_z$), of the PhP electric field distributions.

The validation of the polariton waveguide model can be further manifested in another two ways. First, the near-field profiles at different excitation frequencies were analyzed, which show excellent agreement between the model calculations and experimental measurements (Figure 4e). For the comparison with the FEM-simulated profiles (Figure 4i), one should note that the fringes observed in the FEM calculations are directly launched by the dipole source (see Methods for details), whereas those in the model calculations and experimental measurements are formed by the PhP interferences. Therefore, the FEM calculated fringes exhibit periodicity twice as those from the model and experiment results (Figure 4e and 4i). The dispersion relations of the PhP waveguide modes can then be extracted from these near-field profiles. As shown in Figure 3e, the dispersion relations obtained from the experiments (red dots), model calculations (white line), FEM simulations (black squares), and Fresnel reflection coefficient calculations agree well with each other. The validation can on the other hand be made by applying Fourier transformation of the 2D near-field intensities distributions calculated by the FEM simulations (Figure 4j–4l), which are in accordance with the in-plane dispersion contours calculated by the waveguide model (Figure S5b–5d, Supporting Information).

In conclusion, an analytical waveguide model for describing the polariton propagations in 2D vdW crystals of finite thicknesses was proposed. This model here is able to calculate the dispersion contours, dispersion relations, and localized electromagnetic field distributions, which are important properties of the polaritons in 2D crystals. The efficacy of the model is demonstrated by applying it to calculate the PhP behaviors in the α-$MoO_3$ flake, which reveal that the PhP are waveguide modes. The model results are further verified by numerical FEM simulations and correlative s-SNOM measurements. Our findings can provide a theoretical tool for interpreting the experimental s-SNOM results, therefore help in-depth understand light–matter interactions in the 2D crystals mediated by polaritons.

## METHODS

**Sample preparation.** The α-$MoO_3$ flakes were synthesized by thermal physical deposition method using a tube furnace. The $MoO_3$ powder as source was placed at the center of the quartz tube. The cleaned $SiO_2$ substrate was placed at the low temperature zone. The tube was heated up to 780 °C and then kept at that temperature for 2 hours. Thereafter, α-$MoO_3$ flakes of specific size was selected. The circular holes were fabricated using high-resolution electron beam lithography.

**Near-field optical imaging.** The near-field optical measurements were conducted using a commercially available s-SNOM (NeaSNOM, Neaspec GmbH). To image the PhP in real space, a mid-infrared laser (quantum cascade lasers: www.daylightsolutions.com) with tunable frequencies from 900 $cm^{-1}$ to 1240 $cm^{-1}$ was focused onto the sample through a metal-coated AFM tip (Arrow-IrPt, NanoWorld). The tip was vibrated vertically with a frequency of 280 kHz. The backscattered light from the tip is collected by a mercury cadmium telluride detector (HgCdTe, Kolmar Technologies), and the optical signal was demodulated in a pseudoheterodyne interferometric manner.

**FEM numerical simulations.** The real-space electric filed distributions, Re($E_z$), on the surface of the α-$MoO_3$ flake were calculated using the FEM simulations (COMSOL Multiphysics). As shown in Figure S5a, a vertically-polarized electric dipole source was fixed above the α-$MoO_3$ with a separation of 50 nm. The thicknesses of the air, α-$MoO_3$, and $SiO_2$ layers are 500 nm, 210 nm, and 500 nm, respectively. The permittivities along the three principle axes were calculated according to Equation (S19). The anisotropic dielectric tensor of the α-$MoO_3$ layer is written as

$\begin{bmatrix} \varepsilon_x & 0 & 0 \\ 0 & \varepsilon_y & 0 \\ 0 & 0 & \varepsilon_z \end{bmatrix}$. This tensor was imported into the COMSOL package to solve the Maxwell's equations. The Re($E_z$) was monitored at the surface of the α-$MoO_3$ flake.

ASSOCIATED CONTENT

**Supporting Information**. The following files are available free of charge.

Solution of TE and TM waveguide modes of uniaxial crystal and isotropic two-dimensional waveguide, calculation of isofrequency surfaces of electromagnetic waves propagating inside the homogeneous non-magnetic anisotropic crystal, calculations of the PhP interference patterns, in-plane isofrequency contours of the electromagnetic plane waves in the biaxial α-$MoO_3$ crystal (Figure S1), electromagnetic field distributions in the 2D α-$MoO_3$ flake as functions in-plane propagation angle $\theta$ (Figure S2 and S3), Schematic showing the interference of PhP waves launched by the metallic tip with those reflected by the circular hole (Figure S4), and Fourier transformation of the 2D near-field intensities distributions calculated by the FEM simulations (Figure S5). (PDF)

AUTHOR INFORMATION

**Corresponding Author**

*Huanjun Chen: chenhj8@mail.sysu.edu.cn.

*Shaozhi Deng: stsdsz@mail.sysu.edu.cn.

**Author Contributions**

H.C., S.D., and N.X. conceived the study and supervised the project. F.S. developed the analytical polariton waveguide mode. F.S., H.W., and Z.Z. conducted the experimental measurements and theoretical calculations. Y.K. and R.Z. helped prepare the samples and characterizations. F.S., H.W., Z.Z., H.C., S.D., and N.X. analyzed the data and discussed the results. The manuscript was written through contributions of all authors. All authors have given approval to the final version of the manuscript. §These authors contributed equally.

**Notes**

The authors declare no competing financial interest.

ACKNOWLEDGMENT

This work was financially supported by the National Key Basic Research Program of China (grant nos. 2019YFA0210203), the National Natural Science Foundation of China (grant nos. 91963205, 51702372), Guangdong Basic and Applied Basic Research Foundation (grant nos. 2019A1515011355, 2020A1515011329). H.C. acknowledge the support from Cheung Kong Young Scholars Program. Z.Z. acknowledge the project funded by China Postdoctoral Science Foundation (grant no. 2019M663199).

ABBREVIATIONS

vdW, van der Waals; 2D, two-dimensional; PhP, phonon polariton; FEM, finite element method; TM, transverse magnetic modes; TE, transverse electric modes; s-SNOM, scattering-type scanning near-field optical microscope.

REFERENCES

(1) Lin, X.; Yang, Y.; Rivera, N.; López, J. J.; Shen, Y. C.; Kaminer, I.; Chen, H. S.; Zhang, B. L.; Joannopoulos, J. D.; Soljačić, M. All-angle negative refraction of highly squeezed plasmon and phonon polaritons in graphene–boron nitride heterostructures. *Proc. Natl. Acad. Sci. U.S.A.* **2017**, 114, 6717–6721.

(2) Jiang, J.; Lin, X.; Zhang, B. L. Broadband negative refraction of highly squeezed hyperbolic polaritons in 2D materials. *Research* **2018**, 2532819.

(3) Caldwell, J. D.; Kretinin, A. V.; Chen, Y. G.; Giannini, V.; Fogler, M. M.; Francescato, Y.; Ellis, C. T.; Tischler, J. G.; Woods, C. R.; Giles, A. J.; Hong, M. H.; Watanabe, K.; Taniguchi, T.; Maier, S. A.; Novoselov, K. S. Sub-diffractional volume-confined polaritons in the natural hyperbolic material hexagonal boron nitride. *Nat. Commun.* **2014**, *5*, 5221.

(4) Dai, S.; Ma, Q.; Andersen, T.; Mcleod, A. S.; Fei, Z.; Liu, M. K.; Wagner, M.; Watanabe, K.; Taniguchi, T.; Thiemens, M.; Keilmann, F.; Jarillo-Herrero, P.; Fogler, M. M.; Basov, D. N. Subdiffractional focusing and guiding of polaritonic rays in a natural hyperbolic material. *Nat. Commun.* **2015**, *6*, 6963.

(5) Li, P. N.; Lewin, M.; Kretinin, A. V.; Caldwell, J. D.; Novoselov, Taniguchi, T.; Watanabe, K.; Gaussmann, F.; Taubner, T. Hyperbolic phonon-polaritons in boron nitride for near-field optical imaging and focusing. *Nat. Commun.* **2015**, *6*, 7507.

(6) Kauranen, M.; Zayats, A. V. Nonlinear plasmonics. *Nat. Photonics* **2012**, *6*, 737–748.

(7) Panoiu, N. C.; Sha, W.; Lei, D. Y.; Li, G.-C. Nonlinear optics in plasmonic nanostructures. *J. Opt.* **2018**, *20*, 083001.

(8) Basov, D. N.; Fogler, M. M.; García de Abajo, F. J. Polaritons in van der Waals materials. *Science* **2016**, *354*, aag1992.

(9) Low, T.; Chaves, A.; Caldwell, J. D.; Kumar, A.; Fang, N. X.; Avouris, P.; Heinz, T. F.; Guinea, F.; Martin-Moreno, L.; Koppens, F. Polaritons in layered two-dimensional materials. *Nat. Mater.* **2017**, *16*, 182–194.

(10) Jablan, M.; Buljan, H.; Soljačić, M. Plasmonics in graphene at infrared frequencies. *Phys. Rev. B* **2009**, *80*, 245435.

(11) Koppens, F. H. L.; Chang, D. E.; García de Abajo, F. J. Graphene plasmonics: a platform for strong light–matter interactions. *Nano Lett.* **2011**, *11*, 3370–3377.

(12) Fei, Z.; Andreev, G. O.; Bao, W.; Zhang, L. M.; McLeod, A. S.; Wang, C.; Stewart, M. K.; Zhao, Z.; Dominguez, G.; Thiemens, M.; Fogler, M. M.; Tauber, M. J.; Castro-Neto, A. H.; Lau, C. N.; Keilmann, F.; Basov, D. N. Infrared Nanoscopy of Dirac Plasmons at the Graphene–$SiO_2$ Interface. *Nano Lett.* **2011**, *11*, 4701–4705.

(13) Chen, J.; Badioli, M.; Alonso-González, P.; Thongrattanasiri, S.; Huth, F.; Osmond, J.; Spasenovic, M.; Centeno, A.; Pesquera, A.; Godignon, P.; Elorza, A. Z.; Camara, N.; García de Abajo, F. J.; Hillenbrand, R.; Koppens, F. H. L. Optical nano-imaging of gate-tunable graphene plasmons. *Nature* **2012**, *487*, 77–81.

(14) Fei, Z.; Rodin, A. S.; Andreev, G. O.; Bao, W.; McLeod, A. S.; Wagner, M.; Zhang, L. M.; Zhao, Z.; Thiemens, M.; Dominguez, G.; Fogler, M. M.; Castro-Neto, A. H.; Lau, C. N.; Keilmann, F.; Basov, D. N. Gate-tuning of graphene plasmons revealed by infrared nano-imaging. *Nature* **2012**, *487*, 82–85.

(15) Dai, S.; Fei, Z.; Ma, Q.; Rodin, A. S.; Wagner, M.; McLeod, A. S.; Liu, M. K.; Gannett, W.; Regan, W.; Watanabe, K.; Taniguchi, T.; Thiemens, M.; Dominguez, G.; Castro Neto, A. H.; Zettl, A.; Keilmann, F.; Jarillo-Herrero, P.; Fogler, M. M.; Basov, D. N. Tunable phonon polaritons in atomically thin van der Waals crystals of boron nitride. *Science* **2014**, *343*, 1125–1129.

(16) Caldwell, J. D.; Glembocki, O. J.; Francescato, Y.; Sharac, N.; Giannini, V.; Bezares, F. J.; Long, J. P.; Owrutsky, J. C.; Vurgaftman, I.; Tischler, J. G.; Wheeler, V. D.; Bassim, N.

D.; Shirey, L. M.; Kasica, R.; Maier, S. A. Low-Loss, extreme subdiffraction photon confinement via silicon carbide localized surface phonon polariton resonators. *Nano Lett.* **2013**, *13*, 3690–3697.

(17) Caldwell, J. D.; Aharonovich, I.; Cassabois, G.; Edgar, J. H.; Gil, B.; Basov, D. N. Photonics with hexagonal boron nitride. *Nat. Rev. Mater.* **2019**, *4*, 552–567.

(18) Ma, W.; Alonso-González, P.; Li, S.; Nikitin, A. Y.; Yuan, J.; Martín-Sánchez, J.; Taboada-Gutiérrez, J.; Amenabar, I.; Li, P.; Vélez, S.; Tollan, C.; Dai, Z.; Zhang, Y.; Sriram, S.; Kalantar-Zadeh, K.; Lee, S.-T.; Hillenbrand, R.; Bao, Q. In-plane anisotropic and ultra-low-loss polaritons in a natural van der Waals crystal. *Nature* **2018**, *562*, 557–562.

(19) Zheng, Z.; Chen, J.; Wang, Y.; Wang, X.; Chen, X.; Liu, P.; Xu, J.; Xie, W.; Chen, H.; Deng, S.; Xu, N. Highly confined and tunable hyperbolic phonon polaritons in van der Waals semiconducting transition metal oxides. *Adv. Mater.* **2018**, *30*, 1705318.

(20) Zheng, Z.; Xu, N.; Oscurato, S. L.; Tamagnone, M.; Sun, F.; Jiang, Y.; Ke, Y.; Chen, J.; Huang, W.; Wilson, W. L.; Ambrosio, A.; Deng, S.; Chen. H. A mid-infrared biaxial hyperbolic van der Waals crystal. *Sci. Adv.* **2019**, *5*, eaav8690.

(21) Fei, Z.; Scott, M. E.; Gosztola, D. J.; Foley IV, J. J.; Yan, J.; Mandrus, D. G.; Wen, H.; Zhou, P.; Zhang, D. W.; Sun, Y.; Guest, J. R.; Gray, S. K.; Bao, W.; Wiederrecht, G. P.; Xu, X. Nano-optical imaging of $WSe_2$ waveguide modes revealing light-exciton interactions. *Phys. Rev. B* **2016**, *94*, 081402.

(22) Hu, F.; Luan, Y.; Scott, M. E.; Yan, J.; Mandrus, D. G.; Xu, X.; Fei, Z. *Nat. Photonics* **2017**, 11, 356–360.

(23) Mrejen, M.; Yadgarov, L.; Levanon, A.; Suchowski, H. Transient exciton-polariton dynamics in $WSe_2$ by ultrafast near-field imaging. *Sci. Adv.* **2019**; 5, eaat9618.

(24) Hu, F.; Luan, Y.; Speltz, J.; Zhong, D.; Liu, C. H.; Yan, J.; Mandrus, D. G.; Xu, X.; Fei, Z. Imaging propagative exciton polaritons in atomically thin $WSe_2$ waveguides. *Phys. Rev. B* **2019**, *100*, 121301.

(25) Zheng, Z.; Wang, W.; Ma, T.; Deng, Z.; Ke, Y.; Zhan, R.; Zou, Q.; Ren, W.; Chen, J.; She, J.; Zhang, Y.; Liu, F.; Chen, H.; Deng, S.; Xu, N. Chemically-doped graphene with improved surface plasmon characteristics: an optical near-field study. *Nanoscale* **2016**, *8*, 16621–16630.

(26) Taboada-Gutiérrez, J.; Álvarez-Pérez, G.; Duan, J.; Ma, W.; Crowley, K.; Prieto, I.; Bylinkin, A.; Autore, M. Volkova, H.; Kimura, K.; Kimura, T.; Berger, M.-H.; Li, S.; Bao, Q.; Gao, X. P. A.; Errea, I.; Nikitin, A. Y.; Hillenbrand, R.; Martín-Sánchez, J.; Alonso-González, P. Broad spectral tuning of ultra-low-loss polaritons in a van der Waals crystal by intercalation. *Nat. Mater.* **2020**, doi.org/10.1038/s41563-020-0665-0.

(27) Wunsch, B.; Stauber, T.; Sols, F.; Guinea, F. Dynamical polarization of graphene at finite doping. *New J. Phys.* **2006**, *8*, 318.

(28) Nikitin, A. Y.; Alonso-González, P.; Vélez, S.; Mastel, S.; Centeno, A.; Pesquera, A.; Zurutuza, A.; Casanova, F.; Hueso, L. E.; Koppens, F. H. L.; Hillenbrand, R. A. Real-space mapping of tailored sheet and edge plasmons in graphene nanoresonators. *Nat. Photonics* **2016**, *10*, 239–243.

(29) Álvarez-Ṕerez, G.; Voronin, K. V.; Volkov, V. S.; Alonso-Gonźalez, P.; Nikitin, A. Y. Analytical approximations for the dispersion of electromagnetic modes in slabs of biaxial crystals. *Phys. Rev. B* **2019**, 100, 235408.

(30) Narimanov, E. E. Dyakonov waves in biaxial anisotropic crystals. *Phys. Rev. A* **2018**, *98*, 013818.

(31) Folland, T. G.; Caldwell, J. D. Precise control of infrared polarization. *Nature* **2018**, *562*, 499–501.

(32) Py, M. A.; Schmid, P. E.; Vallin, J. T. Raman scattering and structural properties of $MoO_3$. *IL Nuovo Cimento B* **1977**, *38*, 271–279.

(33) Eda, K. Longitudinal-transverse splitting effects in IR absorption spectra of $MoO_3$. *J. Solid State Chem.* **1991**, *95*, 64–73.

(34) Hillenbrand, R.; Taubner, R.; Keilmann, F. Phonon-enhanced light–matter interaction at the nanometre scale. *Nature* **2002**, *418*, 159–162.

(35) Huber, A.; Ocelic, N.; Kazantsev, D.; Hillenbrand, R. Near-field imaging of mid-infrared surface phonon polariton propagation. *Appl. Phys. Lett.* **2005**, 87, 081103.

TOC: An analytical waveguide model has been developed to quantitatively describe the near-field electromagnetic field distributions, dispersion contours, and dispersion relations of the polariton modes sustained in two-dimensional van der Waals crystals.

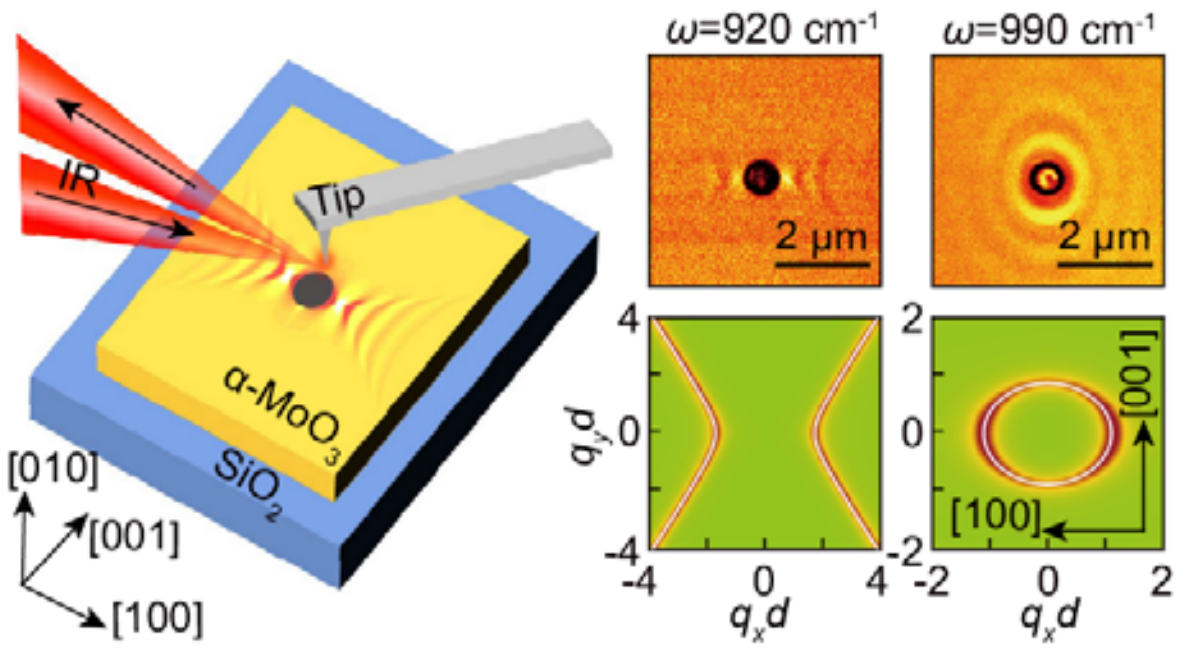